\documentclass[preprint]{iucr}              
\usepackage{bm}
\usepackage{mathtools}
\usepackage{amssymb}
\usepackage{amsmath}
\usepackage{siunitx}
\usepackage[hyphens,spaces,obeyspaces]{url}
\usepackage{color}
\usepackage{siunitx}
\usepackage[hyphens,spaces,obeyspaces]{url}
\usepackage{color}
\usepackage{textgreek}
\usepackage[normalem]{ulem}
\usepackage{makecell}

\newcommand{\inblue}[1]{{\color{black}#1}}

     \journalcode{S}              

\begin{document}                  



\title{X-ray focusing by bent crystals: focal positions as predicted by the crystal lens equation and the dynamical diffraction theory}
%


\cauthor[a]{Jean-Pierre}{Guigay}{guigay@esrf.eu}{address if different from \aff}
\author[a]{Manuel}{Sanchez del Rio}

\aff[a]{European Synchrotron Radiation Facility, 71 Avenue des Martyrs F-38000 Grenoble \country{France}}









\maketitle                        

\begin{synopsis}
A crystal lens equation is deduced to address the location of the focus when monochromatic x-ray radiation encounters a bent crystal. It is extended using dynamical theory of diffraction for Laue symmetrical diffraction. Combination of polychromatic and monochromatic focusing is also discussed. 
\end{synopsis}


\begin{abstract}
The location of the beam focus  when monochromatic x-ray radiation is diffracted by a thin bent crystal is predicted by ``crystal lens equation". We derive this equation in a general form valid for Bragg and Laue geometries. This equation has little utility for diffraction in Laue geometry. The focusing effect in the Laue symmetrical case is discussed using concepts of dynamical theory and an extension of the lens equation is proposed. The existence of polychromatic focusing is considered and the feasibility of matching the polychromatic and monochromatic focal positions is discussed.   \\

\begin{flushright}
{Dedicated to the memory of Claudio Ferrero}
\end{flushright}

\end{abstract}


\section{Introduction}

The use of curved crystals to diffract and focus x-rays comes as a natural extension of the mirror and grating technology for radiation of longer wavelength. Some fundamental concepts, like the Rowland circle, date back to the 19$^\text{th}$ century \cite{rowland1882}.

The fundamental setups using bent crystals to focus X-rays were proposed in the early 1930’s. Some systems use meridional focusing (in the diffraction plane), like i) Johann spectrometer \cite{Johann1931}, using a cylindrically bent crystal,  ii) Johansson spectrometer \cite{Johansson1933} using a ground and cylindrically bent crystal and iii) the Cauchois spectrometer \cite{cauchois1933} in transmission (Laue) geometry. The von Hamos spectrometer \cite{V.Hamos1933} applies sagittal focusing in the plane perpendicular to the diffraction plane.

With the advent of synchrotron radiation, the concepts of ``geometrical focusing" were applied to design instruments such as polychromators for energy-dispersive extended x-ray absorption fine structure (EXAFS) \cite{Tolentino:ms0206}, monochromators with sagittal focusing for bending magnet beamlines \cite{Sparks1980}, or several types of crystal analyzers used at inelastic x-ray scattering beamlines. Bent crystals in transmission or Laue geometry are often employed in beamlines operating at high photon energies. The crystal curvature is used for focusing or collimating the beam in the meridional \cite{Suortti1988,SuorttiShulze} or sagittal \cite{Zhong2001} plane, or just to enlarge the energy bandwidth and improve the luminosity. The crystal bandwidth was optimized and aberrations reduced thanks to the high collimation and small source size of synchrotron beams. Curved crystal monochromators work in off-Rowland condition, whereas crystal analysers for inelastic scattering studies work in the Rowland setting.

A ``Crystal Lens Equation" (CLE) was indeed formulated by \cite{CK} for the focusing properties of a cylindrically bent crystal plate diffracting monochromatic x-rays or neutrons, in Laue (transmission) or Bragg (reflection) geometries. The crystal is bent around an axis perpendicular to the diffraction plane (meridional focusing). This CLE is based on a purely geometric approach in which multiple Bragg scattering (dynamical effects) is neglected. 
The CLE is revisited in Section~\ref{sec:CLE}, in order to correct errors found in \cite{CK} for the Laue geometry. A new formula valid in Bragg and Laue geometry is obtained, using the same geometrical approach as in \cite{CK}.

The CLE has wide applicability in Bragg geometry. However, its use for Laue geometry is limited to very thin crystals, because
it ignores a basic dynamical focusing effect also found in flat crystals,
as described in section~\ref{sec:dynamlicalLaue}. 
The applicability of the lens equation in symmetrical Bragg geometry is discussed in appendix~\ref{sec:BraggGeometry}.
The CLE concerns the focusing of monochromatic radiation, and is in general different from the condition of polychromatic focusing. The particular cases where these two different focusing conditions coincide are discussed in section~\ref{sec:polychromatic}. A final summary is given in section~\ref{sec:summary}.

\section{The crystal lens equation revisited}
\label{sec:CLE}

\begin{figure}
\label{fig:geometries}
\caption{Schematic representation of the different diffraction setups with real or virtual source in Bragg or Laue cases:
a) real source, real focus (red) in Laue case or virtual focus (blue) in Bragg case,
b) real source, virtual focus (red) in Laue case or real focus (blue) in Bragg case,
c) virtual source, real focus (red) in Laue case or virtual focus (blue) in Bragg case,  
d) virtual source, virtual focus (red) in Laue case or real focus (blue) in Bragg case.
$L_0=\overline{SO}$ is the distance source to crystal and $L_h=\overline{OF}$ is the distance crystal to focus.}

\includegraphics[width=0.99\textwidth,trim=4cm 9cm 6cm 9cm,clip=true]{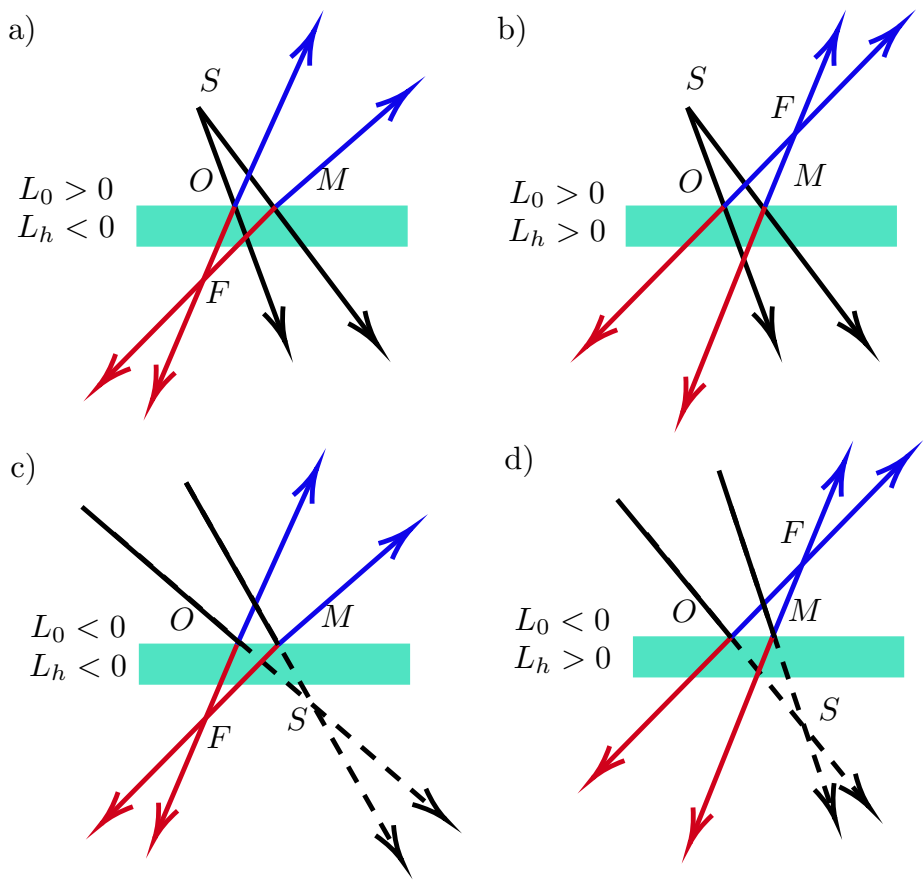}
\end{figure}

\begin{figure}
\label{fig:vectors}
\caption{Schematic view of the relevant parameters in focusing by a bent crystal in Bragg geometry.
}
\includegraphics[width=0.99\textwidth,trim=4cm 6cm 5cm 10cm,clip=true]{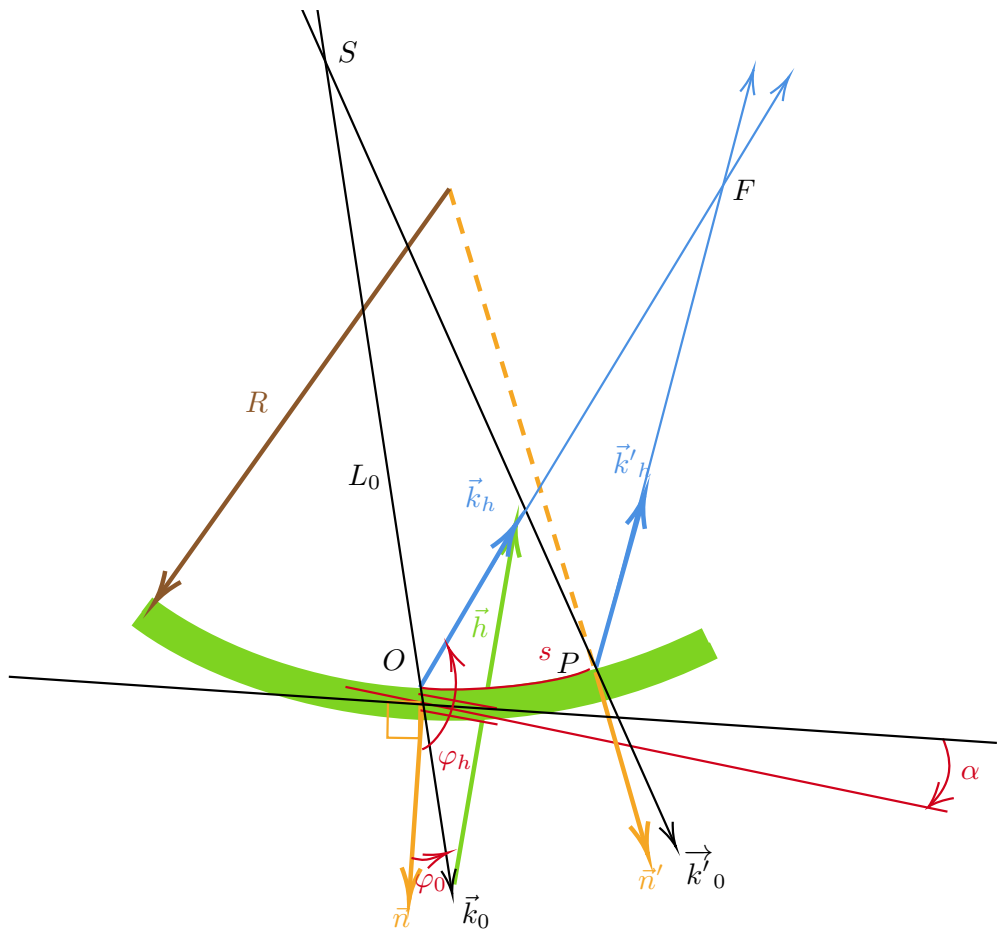}
\end{figure}

The lens equation will be derived in Bragg or Laue geometry, with source $S$ and focus $F$ in real or virtual positions (see Fig.~\ref{fig:geometries}). Consider a monochromatic x-ray or neutron beam from a real or virtual point-source $S$. The origin of coordinates $O$ is chosen at the point of the crystal surface such that the ray $\overline{SO}$, of wavevector  $\vec k_0$, is in \inblue{geometrical} Bragg incidence. \inblue{It gives rise outside the crystal} to a diffracted ray of wavevector $\vec k_h = \vec k_0 + \vec h$, where $\vec h$ is the reciprocal lattice vector in $O$, and \inblue{$|\vec k_h|=|\vec k_0|$} (see Fig.~\ref{fig:vectors}). This is valid in both transmission geometry (Laue) or reflection geometry (Bragg) for both plane and curved crystals\footnote{\inblue{Because of refraction effects, this choice implies that $\overline{SO}$ is, in general, not exactly in the direction of the diffraction profile peak, except for the symmetric Laue case}.}. 

The inward normal to the crystal surface in $O$ is $\vec n$, and $\varphi_0 = (\vec n, \vec k_0)$ is the oriented angle from the vector $\vec n$ to the vector $\vec k_0$. Similarly, $\varphi_h = (\vec n, \vec k_h)$. Without loss of generality $\varphi_0$ is positive; $\theta_B$ is the Bragg angle (always positive).
In the case of symmetric geometry(asymmetry angle $\alpha=0$) we find $\varphi_{0,h}=\pm\theta_B$ in Laue or $\varphi_{0,h}=(\pi/2)\mp\theta_B$ in Bragg. Otherwise, the asymmetry angle $\alpha$ is defined as the angle of rotation of the vector $\vec h$ from its direction in the symmetrical case. 
In Laue case $\varphi_{0,h}=\alpha \pm\theta_B$; in Bragg case $\varphi_{0,h}=\alpha\mp\theta_B+\pi/2$, therefore $2\theta_B=|\varphi_0-\varphi_h|$ in both cases, $2\alpha=\varphi_0+\varphi_h$ in Laue case and $2\alpha=\varphi_0+\varphi_h-\pi$ in Bragg case.

When moving the point of incidence $O$ to $P$ over an arbitrary small distance $s$ along the curved crystal surface (see Fig.~\ref{fig:vectors}), 
$\vec h$ and $\vec n$ are changed into $\vec h'$
and $\vec n'$, respectively. The incident wavevector $\vec k'_{0}$ has the direction of $\overline{SP}$. It is  diffracted into $\vec k'_{h}$. 
The projections of the vectors $\vec k'_{h}$ and $\vec k'_{0}+\vec h'$ on the crystal surface are equal (conservation of the parallel components of wave-vectors).
$\varphi_{0,h}$ are changed into $\varphi'_{0,h}=\varphi_{0,h}+\Delta \varphi_{0,h}$.
Furthermore, in the present case of cylindrical bending of very thin crystal, the surface projection of $\vec h'$ is constant (the angle between $\vec h$ and $\vec n$ is constant).
This implies that $(\sin \varphi_h - \sin \varphi_0)$ is invariant, therefore
\begin{equation}
\label{eq:invariant}
    \Delta \varphi_h \cos\varphi_h = \Delta \varphi_0 \cos\varphi_0.
\end{equation}

The source distance $L_0=\overline{SO}$ is set as positive if the source is on the incidence side of the crystal (real source) or negative if the source is on the other side (virtual source) (see Fig.~\ref{fig:geometries}). The radius of curvature $R_c$ is set as positive if the beam is incident on the concave side of the bent crystal. The focus distance $L_h$ is set as positive if the (real or virtual) focus $F$ is situated on the incidence side on the crystal. With these conventions, $(\vec n,\vec n')=s/R_c$, $\epsilon_0 L_0 = s \cos\varphi_0$,  $\epsilon_h L_h = s |\cos\varphi_h|$,
where $\epsilon_{0,h}$ are the angles between $\vec k_{0,h}$ and $\vec k'_{0,h}$.
Using the relationship
\begin{equation}
    \varphi'_{0,h} = 
    (n',  \vec k'_{0,h}) = 
    (\vec n', \vec n) + (\vec n,\vec k_{0,h}) + (\vec k_{0,h}, \vec k'_{0,h}) = -\frac{s}{R_c} + \varphi_{0,h} + \epsilon_{0,h},
\end{equation}
we obtain
\begin{equation}
\label{eq:angles}
\Delta \varphi_0 =  - \frac{s}{R_c} + s \frac{\cos\varphi_0}{L_0}
\end{equation}
and 
\begin{equation}
\label{eq:angles2}
\Delta \varphi_h = - \frac{s}{R_c} +  s \frac{|\cos\varphi_h|}{L_h}.
\end{equation}

The crystal lens equation valid in both Bragg and Laue cases, is finally obtained by inserting these expressions in equation~(\ref{eq:invariant})

\begin{equation}
\label{eq:CLE}
\frac{|\cos\varphi_h| \cos\varphi_h}{L_h} - \frac{\cos^2\varphi_0}{L_0} = \frac{\cos\varphi_h - \cos\varphi_0}{R_c}.
\end{equation}

In the Laue symmetrical case ($\cos\varphi_h=\cos\varphi_0$) it predicts $L_h=L_0$ (for a real source, the focus is virtual at the same distance as the source) and, in the particular case of $L_0=+\infty$, a plane incident wave is diffracted into a plane wave.

The crystal lens equation~(\ref{eq:CLE}) obtained here is different from the equation given in \cite{CK}\footnote{The CLE given in \cite{CK} is 
$
\cos^2\varphi_0/L_0 + \cos^2\varphi_h/L_h = (\cos\varphi_0 + |\cos\varphi_h|)/R_c$.
We think this is due to mistakes in their calculations, specially sign errors in their equation (9) as compared to our equation~(\ref{eq:angles}). 
}.
Both equations are equivalent for the Bragg case ($\cos\varphi_h<0$), which is also considered by \citeasnoun{snigirevkohn1995}. They are not equivalent in the Laue case.
 
Note that we used in this section the same notation as \cite{CK}, where $R_c$ is positive for a concave surface, used to focus in Bragg case. For the rest of the paper, we also use the notation: $p \leftarrow L_0$, $q \leftarrow -L_h$ $R \leftarrow -R_c$, $\theta_1 \leftarrow \varphi_0$ and $\theta_2 \leftarrow \varphi_h$, which is more convenient for Laue crystals, because real focusing is obtained when the beam coming from a real source is incident on the convex side of the bent crystal (with positive $R$).

Equation~(\ref{eq:CLE}) is obtained here using a geometrical ray optics approach. It can also be deduced from a wave-optics approach as shown in Appendix~\ref{appendix:CLE}.

\section{Dynamical focusing in Laue geometry}
\label{sec:dynamlicalLaue}

The applicability of the CLE for the Laue case is limited to very thin crystals. The dynamical theory (see book \cite{authierbook})
predicts ``new" focal conditions, even for flat Laue crystals.
This is analyzed here in the framework of the Takagi-Taupin equations, hereafter TTE \cite{Takagi1962, Takagi, Taupin, Taupin1967}. 

Section~\ref{sec:influence} deals with the derivation of the ``influence functions" (Green functions) which represent the wavefield generated in the crystal by a point-source on the crystal entrance surface.

In section~\ref{sec:LaueFlat}, the approach to dynamical focusing in the symmetric Laue case
\cite{kushnir, GuigayFerrero2013}
is extended to asymmetric geometry. The effects of anomalous absorption (Borrmann effect) are obtained in parallel. The new concept of ``numerically determined focal length" of a flat crystal, denoted as $q_{dyn}$, is introduced.

In section~\ref{sec:LaueNewCLE}, a lens equation for a bent Laue symmetrical crystal of finite thckness, expressed in terms of $q_{dyn}$ is established. Its predictions are shown to be in agreement with numerical calculations. 

In section~\ref{sec:LaueCompatibilityCLE}, we make the verification that the formulation for the Laue asymmetric case by
\cite{GuigayFerrero2016} is in agreement with the CLE (equation (\ref{eq:CLE})) in the limit of vanishing crystal thickness.

\subsection{Influence function derived from Takagi-Taupin equations}
\label{sec:influence}

The x-ray wavefield inside the crystal is expressed as the sum of two modulated plane waves
\begin{equation}
    \Psi(\vec x) = D_0(\vec x) e^{i \vec k_0 . \vec x} + D_h(\vec x) e^{i \vec k_h . \vec x},
\end{equation}
with slowly varying amplitudes $D_{0,h}(\vec x)$.
The spatial position $\vec x$ is expressed in oblique coordinates $(s_0,s_h)$ along the directions of the $\vec k_0$ and $\vec k_h=\vec k_0 + \vec h$ vectors, which are the in-vacuum wave-vectors of modulus $k=2\pi/\lambda$, where $\lambda$ is x-ray wavelength. $\vec{h}$ is the Bragg diffraction vector of the undeformed crystal. In such conditions, the differential TTE are
\begin{subequations}
\label{eq:TT}
\begin{align}
\frac{\partial D_0}{\partial s_0} =& \frac{ik}{2} \left[ \chi_0 D_0(\vec x)+c \chi_{\bar h} e^{i \vec h . \vec u (\vec x)} D_h(\vec x) \right]; \\
\frac{\partial D_h}{\partial s_h} =& \frac{ik}{2} \left[ \chi_0 D_h(\vec x)+c \chi_{h} e^{-i \vec h . \vec u (\vec x)} D_0(\vec x) \right],
\end{align}
\end{subequations}
where $\chi_0$, $\chi_h$, and $\chi_{\bar h}$ are the Fourier coefficients of order 0, $\vec h$ and $-\vec h$ of the undeformed crystal polarisability. The polarization factor $c$ ($c=1$ for $\sigma$-polarization and $c=\cos2\theta_B$ for $\pi$-polarization) is omitted from now on. 
$\vec u (\vec x)$ is the displacement field of the deformed crystal.
In the case of cylindrical bending we have
\begin{equation}
\label{eq:cylinder}
    \vec h . \vec u = -A s_0 s_h + \phi_1(s_0) - \phi_2(s_h)
\end{equation}
where $A$ and the $\phi_{1,2}$ functions are defined in Appendix~\ref{appendix:Deformation}.
This a 
``constant strain gradient" case \cite{authierbook} meaning that $\partial^2(\vec h . \vec u)/(\partial s_0 \partial s_h)$ is constant. In terms of the functions $G_{0,h}(s_0,s_h)$ defined by
\begin{subequations}
    \label{eq:functionsG}
    \begin{align}
      D_0(s_0,s_h) &= G_0(s_0,s_h) \exp[i\frac{k}{2}\chi_0 (s_0+s_h)-i \phi_2(s_h)]\\
      D_h(s_0,s_h) &= G_h(s_0,s_h) \exp[i\frac{k}{2}\chi_0 (s_0+s_h)-i \phi_1(s_0)+iAs_0s_h],
    \end{align}
\end{subequations}
the TTE have a simpler form
\begin{subequations}
    \label{eq:TTEsimple}
    \begin{align}
      \frac{\partial G_0}{\partial s_0} &= i \frac{k}{2}\chi_{\bar{h}} G_h
      \\
      \frac{\partial G_h}{\partial s_h} &= i \frac{k}{2}\chi_{h} G_0 - i A s_0 G_h.
    \end{align}
\end{subequations}

An incident monochromatic wave of any form can be expressed as a modulated plane wave $D_{inc}(\vec x)\exp(i \vec k_0 . \vec x)$ defining a continuous distribution of coherent elementary point-sources on the crystal surface, according to the general Huyghens principle in optics. The “influence functions” or Green functions, hereafter IF, are the TTE solutions for these point-sources. 
The IF for point-sources of oblique coordinates $(\sigma_0,\sigma_h)$
are derived in \cite{GuigayFerrero2016} by formulating the TTE  as integral equations in the case of an incident amplitude of the form $D_{inc}=\delta(s_h-\sigma_h)$. 
The calculations (see appendix~\ref{appendix:TTEintegral}) result in the diffracted amplitude\footnote{the result for the transmitted amplitude $D_0(s_0,s_h)$ is not necessary for our results and is not presented here, but it is easily obtained using equation~(\ref{eq:kummer}) in (\ref{eq:TTEsimple}).}

\begin{equation}
\label{eq:kummer}
    D_h(s_0,s_h) = \frac{i k }{2} \chi_h e^{(ik/2) \chi_0 (s'_0 + s'_h)} e^{-i \vec h . \vec u (s_0,\sigma_h)} M(\frac{i\Omega}{A},1,iA s'_0 s'_h)
\end{equation}
where the first exponential term stands for the effects of refraction and normal absorption, $s'_{0,h}=s_{0,h}-\sigma_{0,h}$; $\Omega=k^2\chi_h\chi_{\bar{h}}/4$ and the $M$-function is the Kummer function (a confluent hypergeometric function) defined by the convergent infinite series
\begin{equation}
\label{eq:kummerSeries}
    M(a,b,z) = 1 + \frac{a}{b} z + 
    ... + \frac{a(a+1)...(a+n-1)}{n! b (b+1)...(b+n-1)}z^n+...
\end{equation}

This type of TTE solution was already obtained by different methods \cite{Petrashen1974,Katagawa1974,Litzmann1974,Chukhovski1977}.

It is noticeable that the term $\exp[-i\vec h . \vec u (s_0,\sigma_h)]$ in equation~(\ref{eq:kummer}) is the phase shift acquired by scattering at the point of coordinates $(s_0,\sigma_h)$ along the incident ray. We can say that the kinematical (single-scattering) approximation of equation~(\ref{eq:kummer}) is 

\begin{equation}
\label{eq:kummerapprox}
    D_{h,kin}(s_0,s_h) = \frac{i k }{2} \chi_h e^{(ik/2) \chi_0 (s'_0 + s'_h)} e^{-i \vec h . \vec u (s_0,\sigma_h)} 
\end{equation}
and the full multiple scattering is $D_h=D_{h,kin} M$.
          
\subsection{Dynamical focusing and Borrmann effect in a flat, asymmetric, Laue crystal}
\label{sec:LaueFlat}

Dynamical focusing by flat Laue crystals (without bending) was predicted by \citeasnoun{AfanasevKohn1977} and verified experimentally by \cite{Aristov1978,Aristov1980PhysStatSol,Aristov1980}
in the case of symmetrical geometry. The theory was extended to the asymmetric case by \citeasnoun{Kohn2000}. The application of dynamical focusing to high-resolution spectrometry was proposed by \citeasnoun{KohnGorobtsov2013}.

\begin{figure}
\label{fig:laue}
\caption{Schematic representation of the relevant parameters in Laue asymmetrical diffraction.
}
\includegraphics[width=0.99\textwidth,trim=3cm 10cm 5cm 10cm,clip=true]{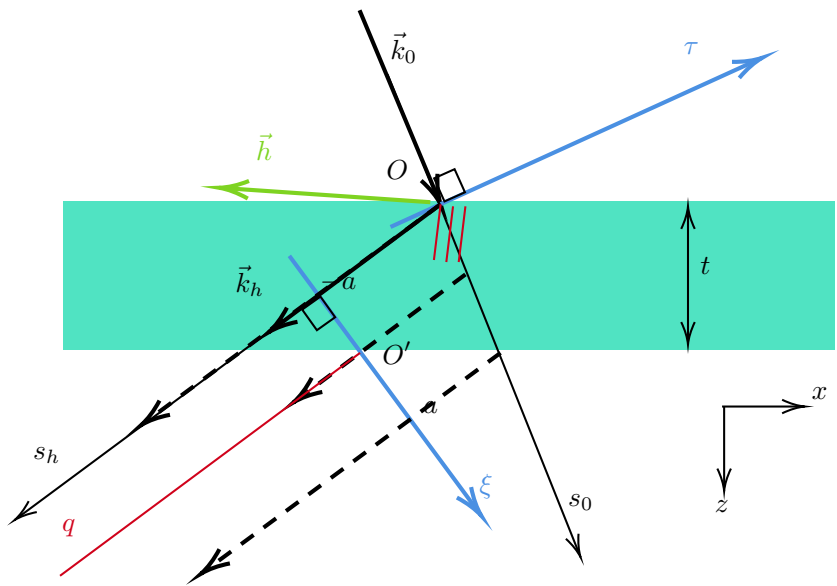}
\end{figure}

The basic case of dynamical focusing is that of a point-source in $O$ ($\sigma_0=\sigma_h=0$) on the crystal entrance surface of the crystal of thickness $t$.
$O'$ is the middle of the basis of the influence region (Borrmann fan) on the exit surface (see Fig.~\ref{fig:laue}).
The amplitude of the diffracted wave along the axis $O'\xi \perp \vec k_h$ is the value of the IF at the point of coordinates 
\begin{equation}
    \label{eq:s0andsh}
    s_0 = \frac{a+\xi}{\sin2\theta_B }  ; \:\: 
    s_h = \gamma\frac{a-\xi}{\sin2\theta_B},
\end{equation}
with $a=t \sin2\theta_B/(2\cos\theta_1)$ and $\gamma=\cos\theta_1/\cos\theta_2$. 

The amplitude $D_h(\xi)$ is zero outside the interval $-a<\xi<a$,
and is proportional to the Bessel function $J_0(k \sqrt{\chi_h \chi_{\bar h} s_0 s_h})=J_0(Z\sqrt{a^2-\xi^2})$ in this interval \cite{kato1961}, with  $Z=k\sqrt{\gamma\chi_h\chi_{\bar h}}/\sin2\theta_B$. In the case $|Z a| \gg 1$ the asymptotic approximation
\begin{equation}
    J_0(Z\sqrt{a^2-\xi^2})\approx \left(\frac{2}{\pi Z \sqrt{a^2-\xi^2}}\right)^{1/2} \cos(Z\sqrt{a^2-\xi^2}-\pi/4)
\end{equation}
can be used in the central region $|\xi|\ll a$ where $\sqrt{a^2-\xi^2} \approx a - \frac{\xi^2}{2a}$.
We thus obtain in this central region the approximation
\begin{equation}
\label{eq:approximatedDiffractedField}
    J_0(Z\sqrt{a^2-\xi^2})\approx \left(\frac{2}{i \pi Z a}\right)^{1/2} \left( e^{iZa-i Z\frac{ \xi^2}{2a}} + i 
    e^{-i Z a+i Z\frac{\xi^2}{2a}} \right),
\end{equation}
where the two exponential terms are related to the two sheets of the dispersion surface. 
The function $\exp(- i Z \xi^2 / (2 a))$ 
represents a converging wave if $\operatorname{Re}(Z)>0$ (divergent if  ($\operatorname{Re} Z <0$). A double, real and virtual, focusing effect is thus expected at opposite distances $\pm q_0$ from the crystal, with
\begin{equation}
\label{eq:q0}
    q_0 = \frac{k a}{|\operatorname{Re}(Z)|}= \frac{a \sin2\theta_B}{|\operatorname{Re}(\sqrt{\gamma\chi_h\chi_{\bar h}})|}
\end{equation}

This equation is present in \cite{Kohn2000, KohnGorobtsov2013} in a different form and from a different point of view. These authors consider a point-source at a finite distance and their equation determines the value of the crystal thickness needed to focus the diffracted wave on the back crystal surface. A noticeable difference is that our equation is expressed in terms of $\chi_h \chi_{\bar h}$ without approximations  concerning the real and imaginary parts of the crystal polarizability. In the works cited above $\operatorname{Re} (\sqrt{\chi_h \chi_{\bar h}})$ is approximated by $|\chi_{hr}|$ or $|\chi_h|$.

\begin{figure}
\label{fig:flatLaue}
\caption{Numerical evaluation of on-axis intensity for a  \SI{250}{\micro\meter} thick flat Si111 crystal ($R=\infty$) with source at the crystal entrance surface ($p=0$) calculated using equation~(\ref{eq:Fresnel}).
a) Simulation for a photon energy of 8.3 keV.
b) Simulation for a photon energy of 17 keV.
Numerical values of these simulations are in Table~\ref{table:example}.
}
\includegraphics[width=1\textwidth]{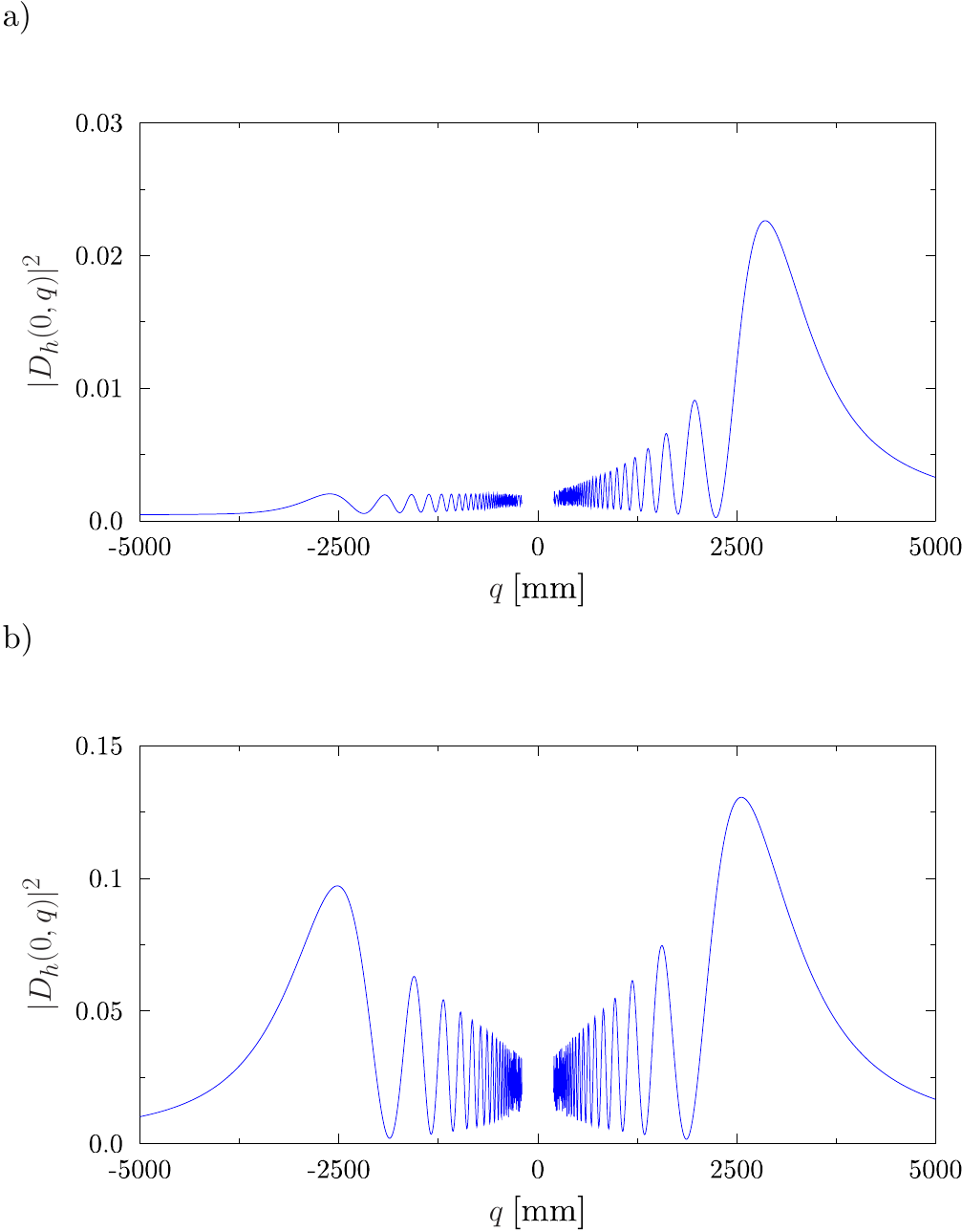}
\end{figure}

The moduli of the two terms in equation~(\ref{eq:approximatedDiffractedField}) are proportional to $\exp(\mp a \text{Im}~(Z))$, respectively. This is the expression of anomalous absorption (Borrmann effect). Two focal positions will be observed for small absorption, but only one for strong absorption, as shown in Fig.~\ref{fig:flatLaue}.

The reflected amplitude at any distance $q$ from the crystal can be calculated numerically, without the approximations used above, by the Fresnel diffraction integral
\begin{equation}
\label{eq:Fresnel}
    D_h(\xi; q) = (\lambda q)^{-1/2} \int_{-a}^a d\xi'  \, e^{i k 
    \frac{(\xi-\xi')^2}{2 q}} 
    J_0(Z\sqrt{a^2-\xi'^2}).
\end{equation}

The ``axial intensity profile" $|D_h(0,q)|^2$ shows in general two strong maxima at distances $q_{1,2}=\pm q_{dyn} < q_0$ (Fig.~\ref{fig:flatLaue}). This difference is a cylindrical aberration effect related to the approximations used to obtain equation (\ref{eq:q0}). The parameter $q_{dyn}$, which depends on the crystal thickness, is the ``dynamical focal length" obtained numerically, thus non-approximated (contrary to $q_0$).
As an example, some numerical values are given in Table~\ref{table:example}. 

\begin{table}
\caption{Parameters for symmetrical Laue silicon crystal in 111 reflection and thickness $t$~= \SI{250}{\micro\meter}.}
\begin{tabular}{llccccc}
 \makecell{Photon \\ energy \\ (keV)}& \makecell{$\theta_B$ \\ (deg)}   & $\chi_0$ & $\chi_h\chi_{\bar h}$ & \makecell{$a$ \\ (\SI{}{\micro\meter})}& \makecell{$q_0$ \\ (mm)} & \makecell{$q_{dyn}$ \\ (mm)} \\
\hline
 8.3  &  13.78 & \makecell{(-14.24 + 0.317 i) 10$^{-6}$} & \makecell{(58.06 - 3.416 i) 10$^{-12}$}  & 59  & 3615  & 2860   \\
 17   &  6.68 & \makecell{(-3.36 + 0.018 i) 10$^{-6}$} & \makecell{(3.20 - 0.046 i) 10$^{-12}$}  & 29  & 3753  & 2535 
\end{tabular}
\label{table:example}
\end{table}

The focusing condition for a source at a finite distance $p$ from the crystal can be obtained by considering that propagation in free-space and propagation in the flat crystal are space-invariant, therefore expressed as convolutions in direct space or simple multiplications in reciprocal space. Therefore, they can be commuted. This allows to merge the free-space propagation before and after the crystal. The focusing condition is therefore
\begin{equation}
    p + q = q_{dyn}.
\end{equation}
On the contrary, propagation through a bent crystal is not space-invariant because the IF is not only dependent on the variables $(s'_0, s'_h)$, but also on the variables $(\sigma_0, \sigma_h)$ because of the factor  
$\exp[-i \vec h . \vec u (s_0, \sigma_h)]$ in 
equation~(\ref{eq:kummer}).

\subsection{A new lens equation for a bent crystal of finite thickness in symmetrical Laue geometry}
\label{sec:LaueNewCLE}

In symmetrical Laue geometry, the factor $\exp(i \chi_0 (s'_0+s'_h))$ in equation~(\ref{eq:kummer}) is constant on the crystal exit surface and will be omitted. Equation~(\ref{eq:kummer}) is (see Appendix~\ref{appendix:TTEintegral})
\begin{equation}
\label{eq:DhSymmetricalLaue}
    D_h(s_0,s_h) = \frac{i k}{2} \chi_h e^{-i \vec h . \vec u(s_0,\sigma_h)}
    J_0(2\sqrt{\Omega s'_0 s'_h}).
\end{equation}

Let us consider the incident amplitude $D_{inc}(\tau)=\exp(i k \tau^2/(2p))$, where $\tau$ is a coordinate along the axis $O\tau$ normal to $\vec k_0$ (see Fig.~\ref{fig:laue}). On the exit surface, using $s_0=(\xi+a)/\sin2\theta_B$ and $\sigma_h=-\tau/\sin2\theta_B$, and the notation $R'=R\cos\theta_B$ we obtain from equations in appendix \ref{appendix:Deformation}, in the case $\alpha=0$

\begin{equation}
    \vec h . \vec u(s_0,\sigma_h) = k \frac{\tau(\tau+a)-\xi(\xi+a)}{ 2R'}.
\end{equation}
Using the integration variable $\eta=\xi-\tau$, the amplitude along the $\xi$-axis is, with omission of $i(k/2)\chi_h$,

\begin{equation}\label{eq:blabla}
    D_h(\xi,0)=\int_{-a}^{+a}\frac{d\eta}{\sqrt{\lambda p}}
    e^{\frac{ik}{2}\left[\frac{(\xi-\eta)^2}{p}+\frac{\eta^2-2\eta\xi-a\eta}{R'}\right]}
    J_0(Z\sqrt{a^2-\eta^2}).
\end{equation}
The wave amplitude at a distance $q$ downstream from the crystal is obtained using a Fresnel diffraction integral similar to equation~(\ref{eq:Fresnel}). We thus have a double integral over $\eta$ and $\xi'$. The $\xi'$ integration is performed analytically \cite{GuigayFerrero2013} and it turns out that
\begin{equation}
\label{eq:Dhpropagated}
    D_h(\xi,q)=
    \frac{e^{i k \frac{\xi^2}{2L}}}{\sqrt{\lambda L}}
    \int_{-a}^{+a} d\eta
    e^{\frac{ik}{2}
    [\frac{\eta^2}{L_e}-\eta(
    \frac{2\xi q_e}{q L_e}+
    \frac{a}{R'}
    )]}
    J_0(Z\sqrt{a^2-\eta^2}),
\end{equation}
where $L=p+q$, $p_e^{-1}=p^{-1}+R'^{-1}$, $q_e^{-1}=q^{-1}-R'^{-1}$ and $L_e=p_e+q_e$. The focal positions are given by $L_{e}=\pm q_{dyn}$.
This can be written as
\begin{equation}
\label{eq:preLaueCLE}
    \frac{R'}{R'-q} - \frac{R'}{R' + p} = \pm \frac{q_{dyn}}{R'}.
\end{equation}
Translating equation~(\ref{eq:preLaueCLE}) in the notation of section~\ref{sec:CLE} ($p \to L_0$, $q \to -L_h$, $R \to -R_c$), we obtain
\begin{equation}
\label{eq:newCLE}
    \frac{1}{L_h-R_c \cos\theta_B} -
    \frac{1}{L_0 - R_c \cos\theta_B} =
    \pm \frac{q_{dyn}}{(R_c \cos\theta_B)^2}.
\end{equation}
If $q_{dyn}$ is set to zero, we obtain $L_h=L_0$, the same result as the lens equation~(\ref{eq:CLE}).
Equation~(\ref{eq:newCLE}) can be considered as a ``modified lens equation" which takes dynamical diffraction effects into account in symmetric Laue geometry.
We do not know an equation like equation~(\ref{eq:newCLE}) for the general case of asymmetrical Laue diffraction. However, numerical simulations can be done to obtain the focal positions \cite{Nesterets,GuigayFerrero2016}.

\begin{figure}
\label{fig:8keV}
\caption{Numerical evaluation of diffracted intensity by a \SI{250}{\micro\meter} thick Si 111 symmetric Laue crystal calculated using equation~(\ref{eq:Dhpropagated}) for a bent (R~= \SI{1}{\meter}) crystal and $p$~= \SI{50}{\meter}. 
a) on-axis intensity for a photon energy of 8.3 keV. 
Inset: transverse profile at the focal distances (maximum values):  
$q_1$~= \SI{651}{\milli\meter} (blue), and
$q_2$~= \SI{1330}{\milli\meter} (red).
b) on-axis intensity for a photon energy of 17 keV.
Inset: transverse profile at the focal distances (maximum values):
$q_1$~= \SI{625}{\milli\meter} (blue), and 
$q_2$~= \SI{1372}{\milli\meter} (red).
}
\includegraphics[width=1\textwidth]{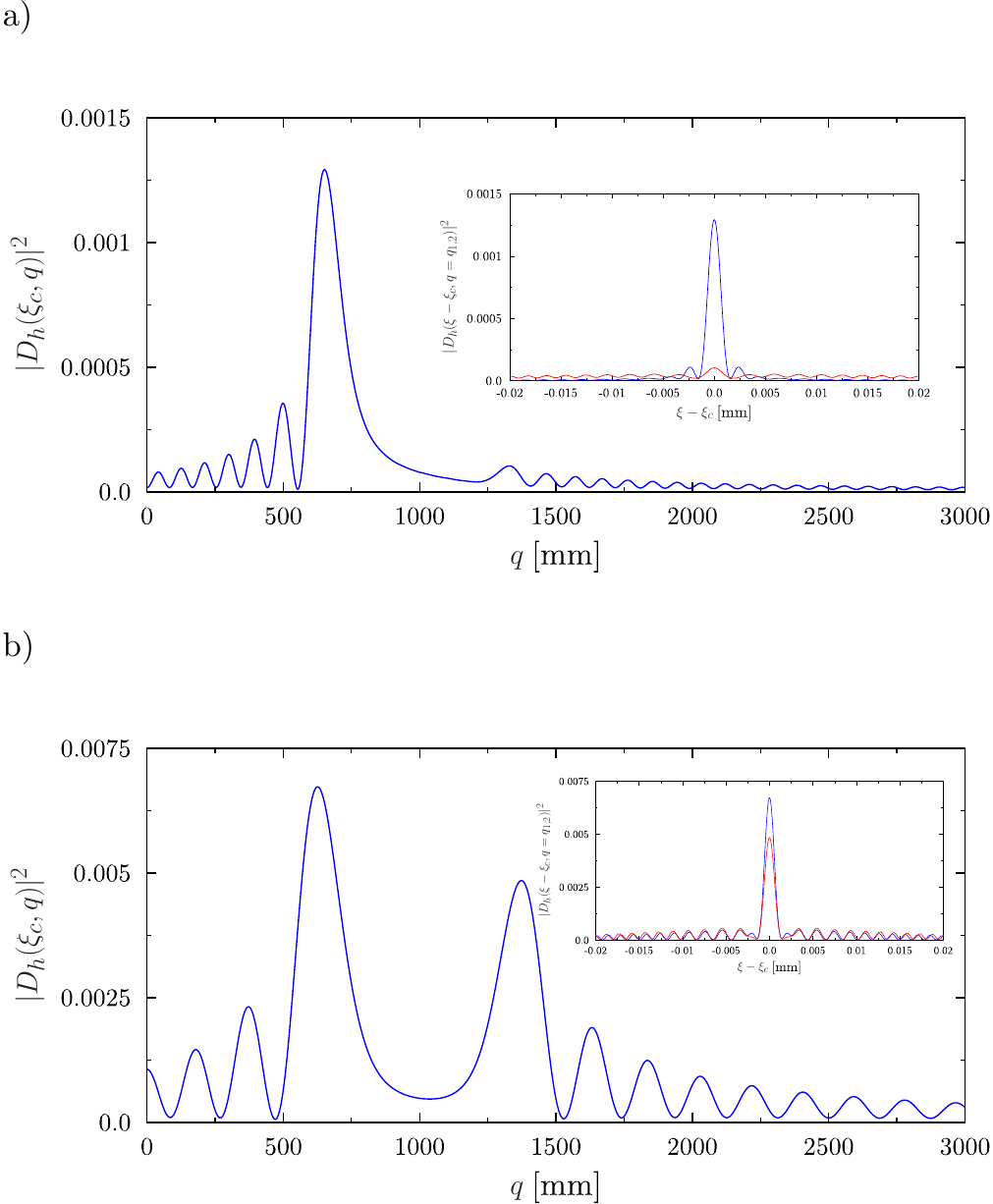}
\end{figure}

Examples of numerical calculations using equation~(\ref{eq:Dhpropagated}) are shown in Fig.~\ref{fig:8keV}, for the case of the 111 reflection of a \SI{250}{\micro\meter} thick cylindrically bent symmetric Laue silicon crystal, with a curvature radius of $R$~= \SI{1}{\meter}, at a source distance $p$~= \SI{30}{\meter} and for x-ray photon energies of 8.3~keV and 17~keV. 

Alternatively, provided that the parameter $q_{dyn}$ has been previously determined numerically by a plot similar to Fig.~\ref{fig:flatLaue}, the focal positions can be given directly by equation~(\ref{eq:newCLE}). The results are in very good agreement with the focal positions obtained obtained numerically in Fig.~\ref{fig:8keV}. An important advantage in using the new CLE is that the same value of $q_{dyn}$ can be used for any value of the radius of curvature and for any value of source distance. 

We are often interested in real focusing ($q>0$) of an incident beam from a very distant real source, for instance in dispersive EXAFS beamlines. 
Suppose $0<R'\le q_{dyn}$. When $p$ increases from zero to infinity, $q_1$ decreases from $q_1=R'q_{dyn}/(q_{dyn}+R')$ to 
$q_1=R' q_{dyn}(q_{dyn}-R')$.
Simultaneously, $q_2$ decreases from $q_2=R'q_{dyn}/(q_{dyn}-R')$ to 
$q_2=R'q_{dyn}(q_{dyn}+R')$.
For very large $p$-values, we have the simple relation $q_1+q_2\approx 2R'$ in good agreement with the numerical results in Fig.~\ref{fig:8keV}.

It can be seen from equation~(\ref{eq:Dhpropagated}) that the intensity function $|D_h(s_0,s_h)|^2$ as a function of $\xi$ is symmetric around $\xi_c=-a q L_e / (2 q_e R')$. This denotes a lateral shift of the intensity profile from its position for the unbent crystal (the axial intensity profiles of Fig.\ref{fig:8keV}a and \ref{fig:8keV}b are actually plotted as a function of $(\xi-\xi_c)$.

\subsection{Semianalytical approach in asymmetric Laue geometry and its CLE limit}
\label{sec:LaueCompatibilityCLE}

The generalization of equation (\ref{eq:blabla}) to asymmetric Laue geometry is \cite{GuigayFerrero2016}
\begin{equation}
\label{eq:unpropagatedkummer}
    D_h(\xi,0) = 
    \int_{-a}^{a} \gamma\frac{d\eta}{\sqrt{\lambda p}}
    e^{i k \gamma^2
    \frac{(\xi-\eta)^2}{2p}+i \phi(\xi,\eta)
    }
     M(\frac{i\Omega}{A},1,i g k \frac{a^2-\eta^2}{R}).
\end{equation}
Here, $\phi(\xi,\eta)$ is calculated from the term $\exp(-i\vec h . \vec u(s_0,\sigma_h))$ in equation~(\ref{eq:kummer}) with $s_0=(a+\xi)/\sin2\theta_B$ and $\sigma_h=\gamma(\xi-\eta)/\sin2\theta_B$, giving
\begin{multline}
    \phi(\xi,\eta) =\frac{k}{2R}[-\mu_2\gamma^2(\xi-\eta)^2
    +a_2\gamma(\eta-\xi) \\
    -\mu_1(a+\xi)^2 
    +a_1(a+\xi)
    -2g(a+\xi)(\xi-\eta)],
\end{multline}
with parameters $\mu_{1,2}$, $a_{1,2}$ and $g$ given in Appendix~\ref{appendix:Deformation}.
The reflected amplitude $D_h(\xi,q)$ at distance $q$ downstream from the crystal is again obtained as in equation~(\ref{eq:Fresnel}), therefore by double integration over $\eta$ and $\xi'$. The $\xi'$-integration can be again performed analytically. The remaining $\eta$-integration involving the Kummer function is carried out numerically \cite{GuigayFerrero2016}. We consider this approach as semi-analytical, in contrast to the approach based on a numerical solution of the TTE \cite{Nesterets}.

It is interesting to study analytically the limit of this semi-analytical formulation in the case of
vanishing crystal thickness ($a\rightarrow0{}$) because the comparison with lens equation~(\ref{eq:CLE}) represents a validity test of the semi-analytical formulation. 
In the limit ($a\rightarrow0{}$),
the Kummer function is equal to unity in equation~(\ref{eq:unpropagatedkummer}), and the integral can be replaced by $2a$ times the integrand evaluated at $\eta=a=0$, therefore

\begin{equation}
\label{eq:14reduced}
    D_h(\xi,0) = \frac{2 a \gamma}{\sqrt{\lambda p}} e^{\frac{i k \xi^2}{2}(\frac{\gamma^2}{p}-\frac{\mu_2\gamma^2+\mu_1+2g}{R})}.
\end{equation}

This is the expression of the amplitude of a cylindrical wave focused at the distance $q$ such that
\begin{equation}
    \frac{1}{q}+\frac{\gamma^2}{p}-\frac{\mu_2\gamma^2+\mu_1+2g}{R}=0. 
\end{equation}
Using the identity
\begin{equation}
\label{eq:appendixIdentity}
    \mu_1+\gamma^2\mu_2+2g=\frac{\cos\theta_2-\cos\theta_1}{\cos^2\theta_2},
\end{equation}
which is derived in Appendix~\ref{appendix:Deformation}, the focusing condition is 
\begin{equation}
    \frac{1}{q}+\frac{\gamma^2}{p}+\frac{\cos\theta_1-\cos\theta_2}{R\cos^2\theta_2}=0,
\end{equation}
or,
\begin{equation}
    \frac{\cos^2\theta_2}{q}+\frac{\cos^2\theta_1}{p}+\frac{\cos\theta_1-\cos\theta_2}{R}=0,
\end{equation}
which is the CLE (equation~(\ref{eq:CLE})) for the Laue case, with the correspondence $p \rightarrow L_0$, $q \rightarrow -L_h$, $R \rightarrow -R_c$, $\theta_1 \rightarrow \varphi_0$ and $\theta_2 \rightarrow \varphi_h$

\section{Polychromatic geometric focusing}
\label{sec:polychromatic}
%

As pointed out by \cite{CK}, the monochromatic focusing condition must not be confused with the polychromatic focusing condition  \cite{handbook,Caciuffo1987,Schulze1998,Martinson,martinson2017}, obtained by varying the wavelength of the reflected rays in order to satisfy the exact Bragg condition on the whole crystal surface.
The equation $\varphi_0+\varphi_h=2\alpha$ in Laue, or $\varphi_0+\varphi_h=2\alpha+\pi$ in Bragg case, implies $\Delta\varphi_0+\Delta\varphi_h=0$. Using equations~(\ref{eq:angles}) and  (\ref{eq:angles2}) we obtain
\begin{equation}
\label{eq:polychromaticfocusing}
\frac{{\cos {\varphi _o}}}{{{L_0}}} + \frac{{\left| {\cos {\varphi _h}} \right|}}{{{L_h}}} = \frac{2}{R_c}.
\end{equation}
Equation~(\ref{eq:polychromaticfocusing}) is usually referred to as the "geometric focusing" condition for bent crystals. It is also applied in the case of flat crystals \cite{sanchezdelrio1994}. Like in equation~(\ref{eq:CLE}), the crystal thickness does not appear in equation~(\ref{eq:polychromaticfocusing}).
The combination of equations (\ref{eq:CLE}) and (\ref{eq:polychromaticfocusing}) gives

\begin{equation}
\label{eq:coincidence}
\frac{\cos\varphi_0}{L_0}(\cos\varphi_h+\cos\varphi_0) = \frac{|\cos\varphi_h|}{L_h}(\cos\varphi_h+\cos\varphi_0),
\end{equation}
which is verified either in the symmetric Bragg case ($\cos\varphi_h+\cos\varphi_0=0$), or if $\cos\varphi_0/L_0=|\cos\varphi_h|/L_h=1/R$, which is the Rowland condition. The Rowland condition is therefore necessary for the coincidence of equations (\ref{eq:CLE}) and (\ref{eq:polychromaticfocusing}) in Laue geometry.
\footnote{This is different from the statement of \cite{CK} that the coincidence is always realised under symmetrical reflection or the Rowland condition.}

A narrow energy band is reflected in Rowland condition, because the angle of incidence on the local reflecting plane does not change along the bent crystal surface.

On synchrotron dispersive EXAFS beamlines, the use of a Bragg symmetric reflection by a bent polychromator at a large distance from the source guarantees the focusing of a broad bandwidth (up to $~1$ keV) on a small spot \cite{Tolentino:ms0206} at a distance close to $L_h=(R_c\sin\theta_B)/2$. 

Laue polychromators are also used in synchrotron beamlines.
In symmetric Laue geometry, condition (\ref{eq:CLE}) should be replaced by equation~(\ref{eq:newCLE}), which is $L_h \approx R_c \cos\theta_B + (R_c \cos\theta_B) ^2 / q_{dyn}$ if the source distance is very large.
Coincidence with (\ref{eq:polychromaticfocusing}) is then obtained if
$R_c=-q_{dyn}/(2\cos\theta_B)$,
which means real focusing at the distance $|L_h|=q_{dyn}/4$ with beam incidence in the crystal convex side ($R_c<0$). 
If $|L_h|$ is fixed, the required conditions are $|R_c| = 2 |L_h| / \cos\theta_B$ and $q_{dyn}=4|L_h|$. The last condition should be fulfilled by choosing the crystal thickness, as in  \cite{Mocella2004,Mocella2008}.

Another polychromatic condition for Laue geometry has been introduced more recently \cite{Martinson, PengQi, PengQi2021}.
The energy components of a polychromatic ray traversing a bent Laue crystal with finite thickness meet the Bragg condition at different positions along the ray path. They are diffracted with different Bragg angles, therefore they exit in different directions, giving raise to a polychromatic focus from a single ray. The ``magic condition", under which single ray focusing and geometric focusing (equation (\ref{eq:polychromaticfocusing})) would coincide, is achieved by the adequate choice of the asymmetry. The magic condition is independent of the crystal thickness \cite{PengQi2021}. We observe that the magic condition (equation (19) in \cite{PengQi2021}) and the modified lens equation (\ref{eq:newCLE}) are both satisfied in the particular case of symmetric Laue geometry in Rowland configuration.   

\section{Conclusions and future perspectives}
\label{sec:summary}

The crystal lens equation (CLE, equation~(\ref{eq:CLE})) based on the conservation of the parallel component of the wavevector in the diffraction process has been revisited. It includes all cases of symmetric and asymmetric Laue and Bragg geometries. It differs from the previous formulation \cite{CK} in the Laue case. However, in Laue geometry, the lens equation 
can be only applied if the crystal is so thin that important effects resulting from the dynamical theory of diffraction, like the focusing of the Borrmann triangle, can be neglected. We derived the modified lens equation (\ref{eq:newCLE}) which overcomes this restriction in the Laue symmetric case. Consistently, it converges to the CLE if the crystal thickness tends to zero. The generic case of arbitrary asymmetry is left for a future investigation.
The fact that dynamic focusing cannot be achieved in Bragg case (see Appendix D) justifies in some way the larger applicability of the CLE in Bragg case.

The application of the CLE (equation~\ref{eq:CLE}) is restricted to monochromatic focusing. Polychromatic focusing, as used in the polychromators of dispersive EXAFS beamlines, happens when the wavelength of the reflected rays changes to exactly match the Bragg angle. This condition is given by a different lens equation (\ref{eq:polychromaticfocusing}). This implies a specular reflection of the rays on the Bragg planes that is, in general, incompatible with the CLE or the results of dynamical theory, except for the Bragg symmetric case. It has been demonstrated that focii predicted by monochromatic and polychromatic focusing conditions coincide if the source is situated on the Rowland circle. Moreover, such coincidence is also true for any source position (off-Rowland) in symmetric Bragg geometry, but not in symmetric Laue geometry. Here, for the Laue symmetric case, both polychromatic and monochromatic focii can match if the modified lens equation~(\ref{eq:newCLE}) is used instead, but requires a particular choice of the crystal thickness.
The additional effect of focusing a polychromatic ray \cite{PengQi2021} gives the ``magic condition"  for Laue focusing, which implies geometric and single scattering. Further studies would be required to match the magic condition (which does not depend on the crystal thickness) with monochromatic focusing. This could be done by optimizing numerically the crystal thickness using the formulation in section~\ref{sec:LaueCompatibilityCLE}.


\bibliography{iucr} 
\bibliographystyle{iucr}

\appendix
\section{Derivation of the Lens Equation from the phase-factor of the Takagi-Taupin equations}
\label{appendix:CLE}

Under a deformation field $\vec u(\vec r)$, the crystal polarizability is taken as 
$\chi(\vec r-\vec u(\vec r))$, where $\chi(\vec r)$ is the
polarizability of the non-deformed crystal. The Fourier components of the electric susceptibility $\chi_{h,\bar h}$ are multiplied by the phase factors $\exp(-  \vec h . \vec u (\vec r))$ and $\exp(\mp i \vec h . \vec u (\vec r))$, respectively, in the TTE.
In the case of a very thin crystal, the ray reflected at position $x$ on the bent crystal surface is simply affected by the phase factor 
\begin{equation}
    \label{eq:phasefactor11}
    e^{-i \vec h . \vec u(x)} = e^{i k (\cos\varphi_h-\cos\varphi_0) \frac{x^2}{2 R_c} },
\end{equation}
which is obtained using $\vec u(x) = -(x^2/(2R_c))\vec n$ and $\vec n . \vec h = \vec n.(\vec k_h - \vec k_0) = k(\cos\varphi_h-\cos\varphi_0)$. 

In the case of the undeformed crystal, the incident amplitude $\exp[i k \tau^2 / (2L_0)]$, along the axis $O\tau$ 
is translated into 
\begin{equation}
    D_{inc}(\xi) = e^{i k \frac{\xi^2}{2L_0}\left(\frac{\cos\varphi_0}{\cos\varphi_h}\right)^2},
\end{equation}
along the axis $O'\xi$ (see Fig.~\ref{fig:laue}).
This is combined with equation (\ref{eq:phasefactor11}) to obtain the amplitude of the Bragg-reflected wave along the axis $O\xi$
\begin{equation}
\label{eq:A4}
    D_h(\xi) = e^{i k
    \frac{\xi^2}{\cos^2\varphi_h}\left(\frac{\cos\varphi_h-\cos\varphi_0}{2R_c} + \frac{\cos^2\varphi_0}{2L_0}\right)},
\end{equation}
corresponding to a real or virtual focus if the phase of this function is negative or positive respectively. 

Using the convention defined in section~\ref{sec:CLE} (see Fig.~\ref{fig:geometries}), a real focus requires $L_h<0$ in Laue case and $L_h>0$ in Bragg. A cylindrical converging wave has then the form $\exp(i k \xi^2 / (2L_h))$ in Laue and $\exp(-i k \xi^2 / (2L_h))$ in Bragg (we arrive to the same result considering a virtual focus). For both Bragg and Laue cases we can write

\begin{equation}
\label{eq:A5}
D_h(\xi) = e^{i k \frac{\xi^2}{2 L_h}\frac{|\cos\varphi_h|}{\cos\varphi_h}}.   
\end{equation}
 Comparing equations~(\ref{eq:A4}) and (\ref{eq:A5}), we finally obtain
\begin{equation}
    \frac{\cos\varphi_h-\cos\varphi_0}{R_c}+
    \frac{\cos^2\varphi_0}{L_0}=\frac{|\cos\varphi_h|\cos\varphi_h}{L_h},
\end{equation}
which is equivalent to the lens equation (\ref{eq:CLE}).

\section{Derivation of the influence functions equation (\ref{eq:kummer}) from the integral form of the Takagi-Taupin equations}
\label{appendix:TTEintegral}

For a point source in position $(\sigma_0,\sigma_h)$ on the crystal surface, the incident amplitude has the form $D_{inc}= \delta(s_h-\sigma_h)$.
The refracted amplitude is $D_{ref}(s_0,s_h)=\exp(i k \chi_0 s'_0/2)\delta(s'_h)$ using $s'_{0,h}=s_{0,h}-\sigma_{0,h}$.
According to equation~(\ref{eq:functionsG}a), this is transformed in $G_{ref}=E \delta(s'_h)$, with
\begin{equation}
\label{eq:appE}
    E =\exp[-i\frac{k}{2}\chi_0(\sigma_0+\sigma_h)+i \phi_2(\sigma_h)].
\end{equation}
Considering $G_{0,h}(s_0,s_h)$  as functions of $s'_0$ and $s'_h$, the TTE (\ref{eq:TTEsimple}) are

\begin{subequations}
\label{eq:TTEappendix}
\begin{align}
    \frac{\partial G_0(s'_0,s'_h)}{\partial s'_0}=&i\frac{k}{2}\chi_{\bar h} G_h(s'_0,s'_h) \\
    \frac{\partial G_h(s'_0,s'_h)}{\partial s'_h}=&i\frac{k}{2}\chi_{h} G_0(s'_0,s'_h) -i A (s'_0+\sigma_0)G_h(s'_0,s'_h).
\end{align}
\end{subequations}
We define the functions $F_{0,h}(s'_0,s'_h)$ such that
\begin{equation}
\label{eq:TTEappendix2}
    G_{0,h}(s'_0,s'_h) = e^{-iA\sigma_0s'_h} F_{0,h}(s'_0,s'_h).
\end{equation}
The equations (\ref{eq:TTEappendix}) are rewritten as
\begin{subequations}
\label{eq:TTEappendix3}
\begin{align}
    \frac{\partial F_0}{\partial s'_0}=&i\frac{k}{2} \chi_{\bar h} F_h \\
    \frac{\partial F_h}{\partial s'_h}=& i\frac{k}{2}\chi_{h}F_0-i A s'_0 F_h,
\end{align}
\end{subequations}
The refracted amplitude is $F_{ref}=E\delta(s'_h)$. 
Equations (\ref{eq:TTEappendix3}) can be written in the form of integral equations:
\begin{subequations}
\label{eq:TTEappendixIntegral}
\begin{align}
    F_0(s'_0,s'_h) =& E \delta(s'_h) +i\frac{k}{2}\chi_{\bar h}\int_0^{s'_0} d\xi_0F_h(\xi_0,s'_h),\\
    F_h(s'_0,s'_h)=& i\frac{k}{2}\chi_{h}\int_0^{s'_h} d\xi_h F_0(s'_0,\xi_h) - i A s'_0 \int_0^{s'_h} d\xi_h F_h(s'_0,\xi_h).
\end{align}
\end{subequations}
We can combine equations (\ref{eq:TTEappendixIntegral}) into a single integral equation for $F_h$ only
\begin{equation}
    \label{eq:TTEappendixSingleIntegral}
    F_h(s'_0,s'_h) = i\frac{k}{2}\chi_h E - \Omega \int_0^{s'_h} d \xi_h \int_0^{s'_0} d\xi_0 F_h(\xi_0,\xi_h) - i A s'_0 \int_0^{s'_h} d\xi_h F_h(s'_0,\xi_h),
\end{equation}
where $\Omega=k^2\chi_h\chi_{\bar h}/4$ is used. By iteration starting from $F_h=i\frac{k}{2}\chi_hE$, we obtain

\begin{align}
    \label{eq:TTEappendixSeries}
    F_h(s'_0,s'_h) = i\frac{k}{2}\chi_h E [ 1 - 
    (\Omega+i A) s'_0s'_h + ...+
    \nonumber \\
    (\Omega+iA)(\Omega+2iA)...(\Omega+niA)
    \frac{(-s'_0 s'_h)^n}{n!n!}
    +...]
\end{align}

According to the definition of the Kummer function (equation (\ref{eq:kummerSeries})), the series in brackets is equal to $M(\frac{\Omega}{iA}+1, 1, -iA s'_0 s'_h)$. Using the known relation
$M(a,b,z)=e^z M(b-a,b,-z)$, together with equations (\ref{eq:appE}) and (\ref{eq:TTEappendix2}) we obtain

\begin{equation}
    G_h = i\frac{k}{2}\chi_h \exp\left[-i \frac{k}{2} \chi_0 (\sigma_0+\sigma_h)+ i \phi_2(\sigma_h)-i A \sigma_0 s'_h - i A s'_0 s'_h\right] M(i\frac{\Omega}{A},1,i A s'_0 s'_h)
\end{equation}

Using equation (9b) and $s_0s_h-\sigma_0 s'_h - s'_0 s'_h=s_0 s_h - s_0 s'_h = s_0 \sigma_h$, we get
\begin{equation}
    D_h = i \frac{k}{2} \chi_h \exp\left[  i \frac{k}{2} \chi_0 (s'_0+s'_h) - i\phi_1(s_0) + i \phi_2(\sigma_h) - iA s_0 \sigma_h \right] M(i\frac{\Omega}{A},1,i A s'_0 s'_h), 
\end{equation}
which, including equation (\ref{eq:cylinder}), is equation (\ref{eq:kummer}).

In the symmetric Laue case, for which $A=0$, we obtain from equation~(\ref{eq:TTEappendixSeries})

\begin{equation}
    F_h = i \frac{k}{2} \chi_h E J_0(2 \sqrt{\Omega s'_0 s'_h}),
\end{equation}
consequently, equation (\ref{eq:kummer}) becomes
\begin{equation}
    D_h = i \frac{k}{2} \chi_h \exp\left[  i \frac{k}{2} \chi_0 (s'_0+s'_h) - i \vec h . \vec u(s_0,\sigma_h) \right] J_0(2\sqrt{\Omega s'_0 s'_h}).
\end{equation}

In the opposite case when $A >> \Omega$, an approximated solution could be obtained by considering the simplified integral equation
\begin{equation}
        F_h(s'_0,s'_h) = i\frac{k}{2}\chi_h E - i A s'_0 \int_0^{s'_h} d\xi_h F_h(s'_0,\xi_h),
\end{equation}
with solution $F_h(s'_0,s'_h) = i\frac{k}{2}\chi_h E \exp[-i A s'_0 s'_h]$ from which equation (\ref{eq:kummerapprox}) of section \ref{sec:dynamlicalLaue} is obtained.

\section{Expression of the phase factor in Laue geomety and derivation of equation~(\ref{eq:appendixIdentity})}
\label{appendix:Deformation}.

The components of the displacement field, in the case of meridional bending of radius $R$ are \cite{Nesterets}:
\begin{equation}
    u_x = -\frac{x(z-t/2)}{R}; \, u_z=\frac{x^2+\rho(z-t/2)^2}{2R},
\end{equation}
with $\rho=\nu/(1-\nu)$, and $\nu$ the Poisson ratio. 
Note that $h_x=k(\sin\theta_2-\sin\theta_1)$, $h_z=k(\cos\theta_2-\cos\theta_1)$.
In terms of the oblique coordinates $(s_0,s_h)$ along $\vec k_{0,h}$, such that $z=s_0\cos\theta_1 + s_h \cos\theta_2$ and $x=s_0 \sin\theta_1+s_h\sin\theta_2$ , 
it is found, by lengthy but simple calculations and with omission of a constant term, that $\vec h.\vec u=-A s_0 s_h + \phi_1(s_0) -\phi_2(s_h)$,  
with the following definitions
\begin{equation}
    A = -(2 k \sin\theta_B /R)\sin\alpha[1+(1+\rho)\cos\theta_1\cos\theta_2]
\end{equation}
\begin{align}
    \phi_1(s_0) &= \frac{k}{2R}[\mu_1(s_o\sin2\theta_B)^2-a_1 s_0\sin2\theta_B] \nonumber \\
    \phi_2(s_h) &= -\frac{k}{2R}[\mu_2(s_h\sin2\theta_B)^2-a_2 s_h\sin2\theta_B],
\end{align}
where $\gamma=\cos\theta_1/\cos\theta_2$, $\theta_{1,2}=\alpha\pm \theta_B$,
\begin{align}
   g &= \frac{A \gamma R}{k \sin^2 2\theta_B} = -\sin\alpha\frac{\gamma +(1+\rho)\cos^2\theta_1}{\sin2\theta_B\cos\theta_B}, \nonumber \\
   \mu_{1,2} &=\frac{\sin\alpha(\sin^2\theta_{1,2}+\rho\cos^2\theta_{1,2})+\cos\alpha\sin2\theta_{1,2}}{\sin2\theta_B\cos\theta_B}, \nonumber \\
   a_{1,2} &=t\frac{\cos\alpha\sin\theta_{1,2}+\rho\sin\alpha\cos\theta_{1,2}}{\cos\theta_B}. \nonumber
\end{align}

The parameter $\rho$ is eliminated in the following expressions 
\begin{align}
    \mu_1+g=&\frac{\cos\alpha\sin2\theta_1-\sin\alpha(\gamma+\cos2\theta_1)}{\sin2\theta_B\cos\theta_B},
    \\
    \gamma^2\mu_2+g=&\frac{\gamma^2\cos\alpha\sin2\theta_2-\sin\alpha(\gamma+\gamma^2\cos2\theta_2)}{\sin2\theta_B\cos\theta_B}.
\end{align}

Using
\begin{equation}
    \mu_1 + \gamma^2 \mu_2 + 2 g = \frac{
    \sin(\alpha+2 \theta_B)\cos^2\theta_2 + \sin(\alpha-2\theta_B) \cos^2\theta_1 - 2 \sin\alpha \cos\theta_1 \cos\theta_2
    }{
    \sin2\theta_B \cos\theta_B \cos^2\theta_2
    },
\end{equation}
by some cumbersome algebraic manipulations, the numerator of this last expression is reduced to $(\sin\alpha \sin^22\theta_B)$ from where it is easy to get equation (\ref{eq:appendixIdentity}).

\section{Relevance of the lens equation in the symmetric Bragg case}
\label{sec:BraggGeometry}

Let us consider the case of a flat non-absorbing crystal plate (without bending), in symmetrical Bragg geometry. The fact that experimental results and also numerical calculations \cite{Honkanen2018} of Bragg diffraction with plane crystals do not show any focusing effect (contrary to Laue case), can be loosely explained by the following intuitive approach. Consider that any geometrical ray emitted from a real distant point-source produces a reflected ray at the point of incidence on the crystal surface, with a reflectivity coefficient $r$ equal to the complex reflectivity of the incident plane wave having the same glancing angle of incidence $\theta=\theta_B+\Delta\theta$ as the geometrical ray under consideration
\begin{equation}
\label{eq:braggDiffProfile}
    r(\Delta\theta) = \sqrt{1-\left(\frac{\Delta\theta\sin2\theta_B}{|\chi_h|}\right)^2} + i \frac{\Delta\theta\sin2\theta_B}{|\chi_h|} =
    e^{i \arcsin{\frac{\Delta\theta \sin2\theta_B}{ |\chi_h|}}}.
\end{equation}

Note that $|r(\Delta\theta)|^2$ is the usual diffraction profile. Taking the origin of coordinates at the point corresponding to $\theta=\theta_B$, the reflected wave-amplitude along an axis $O\xi$ situated in the diffraction plane and perpendicular to the reflected direction, at negligible distance from the crystal, may be approximated by setting  $\Delta\theta=\xi/p$ in equation~(\ref{eq:braggDiffProfile}):
\begin{equation}
    D(\xi) = e^{i \arcsin{\frac{\xi \sin2\theta}{p |\chi_h|}}}.
\end{equation}

No focusing effect is expected from this amplitude distribution, because the phase function $\arcsin(\xi \sin2\theta/ (p |\chi_h|))$ is an odd function of $\xi$, thus it does not have a second-order term characteristic of a focusing effect. The first-order term produces a lateral shift of the image. There is no equivalent to the dynamical focusing length $q_{dyn}$ introduced in the Laue case. The reflected beam is indeed divergent, as in the case of a usual mirror.

The focusing properties of cylindrically bent crystals in symmetric Bragg geometry were simulated by \cite{sutter2010}, using a finite-difference method, and by \cite{honkanen2017, Honkanen2018}, using a finite-element method for numerical solution of the TTE. The obtained phase distribution of the reflected wavefront shows a parabolic shape, with concavity inversion as compared to the parabolic phase distribution of the incident wavefront. This is a clear indication of a single real focusing effect, which is indeed confirmed by simulating the reflected wave propagation. The obtained focusing distances are indeed in good agreement with the CLE which is $L_0^{-1}+L_h^{-1}=2/(R_c \sin\theta_B)$ in this case.

\end{document}